\documentclass[sigplan,10pt]{acmart}

\renewcommand\footnotetextcopyrightpermission[1]{}
\usepackage{xspace}

\usepackage{amsmath} % For \text, if needed for other parts of a document
\usepackage{algorithm}
\usepackage[noend]{algpseudocode} % Using [noend] for a more compact look by omitting \EndIf etc.

\usepackage{listings}

\AtBeginDocument{%
  }

\newcommand{\oursys}{Strata\xspace}

\begin{document}
\settopmatter{printacmref=false}
\pagestyle{plain}

\title{\oursys: Hierarchical Context Caching for Long \\ Context Language Model Serving}

\author{Zhiqiang Xie}
\affiliation{
  \institution{Stanford University \& NVIDIA}
  \country{}
  % \city{Stanford}\state{CA}\country{USA}
  }
\email{xiezhq@stanford.edu}

\author{Ziyi Xu}
\affiliation{
  \institution{Shanghai Jiao Tong University}
  % \city{Shanghai}\country{China}
  \country{}
  }
\email{xzy2022@sjtu.edu.cn}

\author{Mark Zhao}
\affiliation{
  \institution{University of Colorado Boulder}
  % \city{Boulder}\state{CO}\country{USA}
  \country{}
}
\email{myzhao@colorado.edu}

\author{Yuwei An}
\affiliation{
  \institution{Carnegie Mellon University}
  % \city{Pittsburgh}\state{PA}\country{USA}
  \country{}
  }
\email{ayw.sirius19@gmail.com}

\author{Vikram Sharma Mailthody}
\affiliation{
  \institution{NVIDIA}
  % \city{Santa Clara}\state{CA}\country{USA}
  \country{}
  }
\email{vmailthody@nvidia.com}

\author{Scott Mahlke}
\affiliation{
  \institution{NVIDIA \& University of Michigan}
  % \city{Santa Clara}\state{CA}\country{USA}
  \country{}
  }
\email{mahlke@umich.edu}

\author{Michael Garland}
\affiliation{
  \institution{NVIDIA}
  % \city{Santa Clara}\state{CA}\country{USA}
  \country{}
}
\email{mgarland@nvidia.com}

\author{Christos Kozyrakis}
\affiliation{
  \institution{NVIDIA \& Stanford University}
  % \city{Santa Clara}\state{CA}\country{USA}
  \country{}
  }
\email{kozyraki@stanford.edu}

\begin{abstract}
Large Language Models (LLMs) with expanding context windows face significant performance hurdles.
While caching key–value (KV) states is critical for avoiding redundant computation, the storage footprint of long-context caches quickly exceeds GPU memory capacity, forcing production systems to adopt hierarchical caching across memory hierarchies. However, transferring large cached contexts back to the GPU introduces severe performance bottlenecks: fragmented I/O from paged layouts prevents full bandwidth utilization, and existing schedulers fail to account for cache-loading delays, leaving systems loading-bound rather than compute-bound.

We present \oursys, a hierarchical context caching framework designed for efficient long-context LLM serving. \oursys introduces GPU-assisted I/O to combat KV cache fragmentation, decoupling GPU and CPU memory layouts and employs cache-aware request scheduling to balance compute with I/O latency and overlapping unavoidable stalls with complementary tasks. 
Built on SGLang and deployed in production, \oursys achieves up to $5\times$ lower Time-To-First-Token (TTFT) compared to vLLM + LMCache and $3.75\times$ speedup over NVIDIA TensorRT-LLM on long-context benchmarks, without degrading short-context performance.

\end{abstract}

\maketitle

\section{Introduction}
Large Language Models (LLMs) represent a significant advancement in machine learning, achieving remarkable proficiency in understanding and generating natural language. Their adoption is now widespread, and they are rapidly evolving towards more robust problem-solving capabilities.
A prominent trend in this evolution is the expansion of context windows, allowing LLMs to parse longer input prompts.
This has enabled an important set of applications that require understanding large amounts of text, including coding assistants, retrieval-augmented generation (RAG), document analysis, and conversational AI agents.
Leading models, including Google's Gemini series~\cite{google2023gemini} and the Qwen series~\cite{qwen2.5-1m}, already support context windows of up to one million tokens, with expectations of two million tokens emerging soon.
Other frontier models like DeepSeek-V3~\cite{deepseekv3}, Llama3.1 and Llama-4 series~\cite{llama3.1,llama4}, and Anthropic's Claude 3.5~\cite{claude3.5} also offer substantial context lengths, typically in the range of 128K to 200K tokens.

While long contexts enable new capabilities such as multi-turn conversations, RAG, and document-centric tasks (e.g., querying books, manuals, or scripts), recomputing them from scratch is prohibitively expensive. 
Caching previously computed key–value (KV) states offers a practical solution, as these prefixes and sources are frequently reused across applications. This technique, often referred to as context or prefix caching~\cite{neurips24:zheng_sglang, deepseek_cache,prompt_caching,google_context,anthropic_prompt_caching_docs}, avoids redundant prefill computation and significantly reduces response latency. 
However, the storage footprint of cached KV states is substantial. For example, 40 GB of GPU High-Bandwidth Memory (HBM) can only hold roughly 0.3M tokens for Llama-8B, which can be quickly consumed by a handful of documents or hundreds of conversation turns. As a result, production systems adopt hierarchical caching, storing KV states in CPU memory~\cite{eurosys25:yu_pensieve}, local SSDs~\cite{atc24:gao_cachedattention}, or even remote memory pools~\cite{arxiv24:hu_memserve, fast25:qin_mooncake} to extend capacity and preserve reuse benefits.

\begin{figure}[t]
    \centering
    \includegraphics[width=\linewidth]{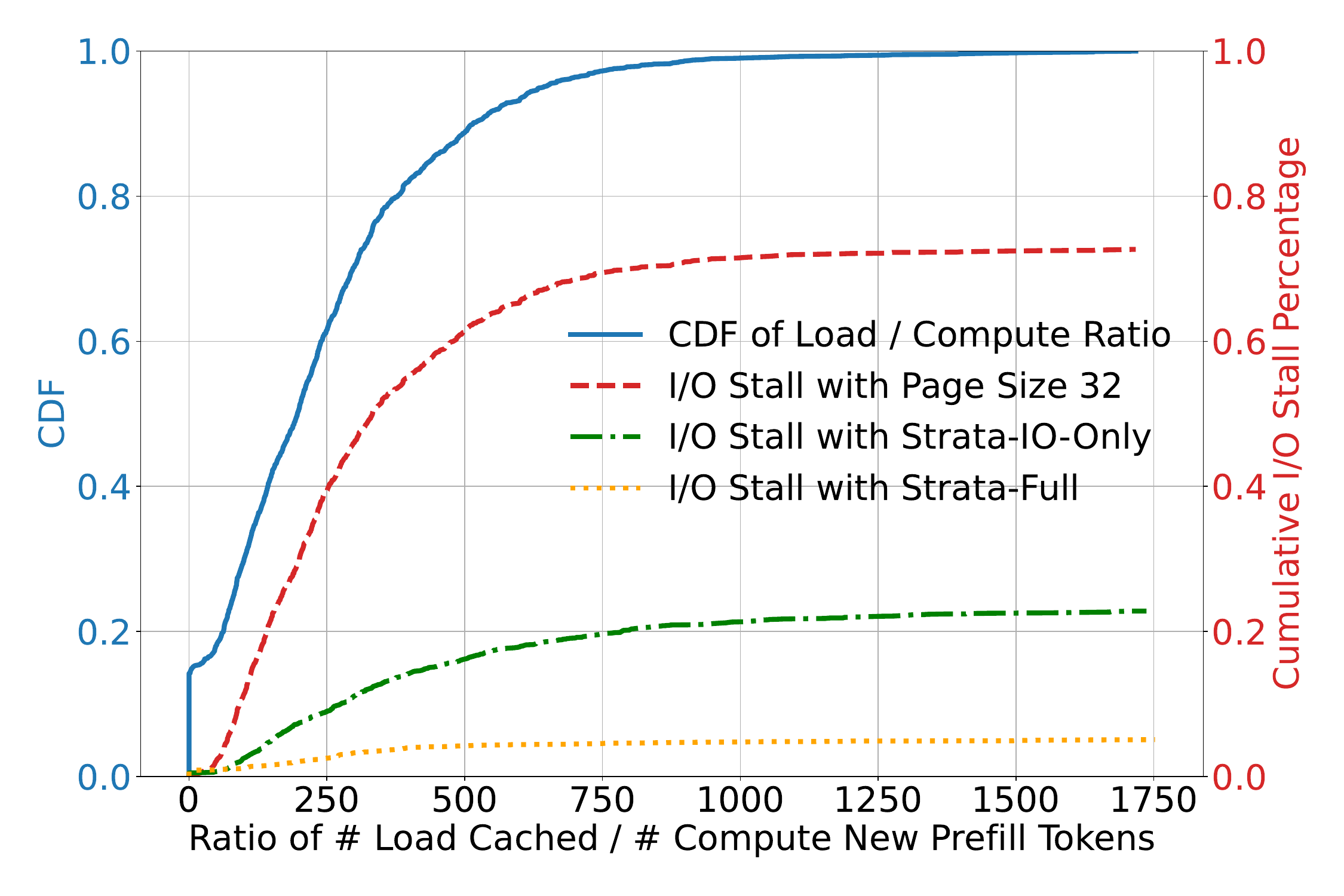}
    \caption{
    Benchmark profile for Qwen2.5-14B on the LooGLE dataset. The x-axis shows the Load / Compute Ratio (tokens loaded from CPU memory relative to new input tokens)  per prefill batch. The right axis displays the I/O stall percentage, representing the amount of prefill execution time attributed to I/O stall. See \S\ref{eval:depth} for full benchmark details.
    }
    \label{fig:stall}
\end{figure}

However, transferring large cached contexts back to the GPU introduces a major performance bottleneck. Bulk KV transfers often cause memory stall, directly inflating TTFT and degrading throughput.
Figure~\ref{fig:stall} illustrates this effect: when serving the LooGLE dataset~\cite{loogle} with SGLang offloading KV caches to CPU memory, configured in line with standard practices reported in prior work~\cite{atc24:gao_cachedattention}, 74\% of prefill time is blocked on KV transfers (the red curve), resulting in up to a $4\times$ throughput reduction. 
In these cases, I/O delays rather than compute become the dominant limiting factor. This inefficiency arises from two main sources.

First, as context lengths grow, the sheer volume of KV cache data requiring transfer between memory tiers (e.g., CPU memory to GPU HBM) increases substantially.
However, current systems achieve only a fraction of the maximum hardware bandwidth between memory tiers.
As we further explore in \S\ref{motivation:bandwidth}, this is because current systems adopt PagedAttention~\cite{sosp23:kwon_vllm} to reduce GPU memory fragmentation.
However, paging causes \textit{data} fragmentation, as the KV cache for a given sequence is spread across multiple non-contiguous pages.
This leads to small data transfers, sometimes only a few kilobytes, which fail to saturate PCIe bandwidth.

In addition to the inefficient I/O, current schedulers fail to account for the fact that loading cached context itself can become a bottleneck.
Specifically, existing systems~\cite{neurips24:zheng_sglang, sosp23:kwon_vllm, website:tensorrt} assumes that the computation needed to prefill new tokens is sufficient to hide the latency of loading historical KV cache from slower memory tiers.
However, as context lengths grow, this assumption no longer holds: caching loading time can exceed the compute time needed for prefill, leaving the system loading-bound rather than compute-bound.
Figure~\ref{fig:stall} highlights this effect. Even with our optimized I/O mechanism presented in \S\ref{sec:design_io}, which removes the overhead of small page transfers (the green line), up to 24\% of prefill execution time remains stalled on cache loading.
Schedulers that ignore these I/O-bound characteristics generate imbalanced batches, unable to effectively hide cache-loading delays.

To address the critical bottlenecks identified above, we propose \oursys, a hierarchical context caching framework designed for long context language model serving, without performance degradation in small context scenarios.
\oursys introduces a novel I/O mechanism to enable more efficient data transfer among GPU HBM, CPU memory, and disk storage.
Specifically, \oursys employs GPU-assisted data transfer to combat KV cache fragmentation and decouples the GPU's memory layout from that of other memory tiers.
Furthermore, \oursys reduces long-context overheads through cache-aware request scheduling. It constructs balanced batches that pair sufficient prefill computation to cover I/O latency, and, when cache loading stalls are unavoidable, schedules insert useful complementary tasks (e.g., decoding batches) to fully utilize available compute resources.
Together, these techniques ensure that scheduling remains efficient even under highly variable latency budgets.

Our implementation of \oursys builds upon SGLang~\cite{neurips24:zheng_sglang}, a widely adopted open-source framework for LLM serving. Our system also has been deployed in production environments at a leading AI company.
We conducted a comprehensive evaluation using popular long-context benchmarks, testing across a range of models and representative hardware platforms.
The results demonstrate that \oursys outperforms vLLM + LMCache~\cite{website:lmcache}, a state-of-the-art open-source hierarchical context caching solution on TTFT, by up to $5\times$, and NVIDIA's TensorRT-LLM, a highly optimized serving engine, by up to $3.75\times$ on these demanding workloads, without performance degradation on short-context scenarios.
\section{Background}

\subsection{Long Context LLM Inference}
LLM inference operates in two phases: \emph{prefill} and \emph{decode}. 
During prefill, the model typically processes both (i) new tokens from the user query and (ii) context tokens, drawn from sources such as documents or prior interactions. 
The intermediate outputs of this step, known as KV caches, are critical for efficiency, as they eliminate the prohibitive cost of recomputation. 
In the subsequent decode phase, the model generates tokens autoregressively, continually reusing and extending the KV cache. 
Thus, efficient cache management is essential to sustaining high-performance serving.

\subsection{Memory Management of KV Cache}
Inspired by virtual memory, PagedAttention~\cite{sosp23:kwon_vllm} avoids reserving large contiguous blocks for KV caches by using dynamic, page-based allocation. 
The cache is partitioned into small fixed-size pages that preserve logical sequence order but can be placed non-contiguously in memory, improving utilization. 
Typical page sizes are small—e.g., 32, 16, and 1 tokens in TensorRT-LLM, vLLM, and SGLang—where each token may span from tens of kilobytes to several megabytes. 
While such fine granularity is manageable for compute kernels, it poses serious efficiency challenges for data movement across memory tiers, as we will discuss in \S\ref{motivation:bandwidth}.

\subsection{Context Caching in LLM Serving}
Beyond intra-request reuse, systems exploit \emph{context caching} across requests by identifying common prefixes using structures like prefix trees or hash maps~\cite{fast25:qin_mooncake, sosp23:kwon_vllm}, widely adopted by providers such as OpenAI~\cite{prompt_caching} and Google~\cite{google_context}. 
To extend capacity, caches are stored in slower tiers such as CPU memory~\cite{eurosys25:yu_pensieve, ragcache, mlsys24:gim_promptcache}, distributed memory pools~\cite{arxiv24:hu_memserve, sigcomm25:liu_cachegen, fast25:qin_mooncake}, or even disk~\cite{atc24:gao_cachedattention, asplos25:flashgen, website:lmcache}. Recent systems, e.g., CachedAttention~\cite{atc24:gao_cachedattention}, overlap cache loading with computation on a layer-by-layer basis to minimize stalls, while asynchronously backing up newly generated caches to lower tiers.

\section{Challenges of Long Context Caching}\label{sec:motivation}

This paper addresses the challenge of managing large context caches for long-context (i.e., prefill-dominated) workloads.
While this is not the only LLM scenario (i.e., short context, long generation, or single-turn workloads exist), long-context workloads represent a significant set of important, real-world workloads~\cite{loogle, kovcisky2018narrativeqa}.
We next explore the systems challenges that arise in long-context workloads.

\begin{figure}[t]
    \centering
    \includegraphics[width=.95\linewidth]{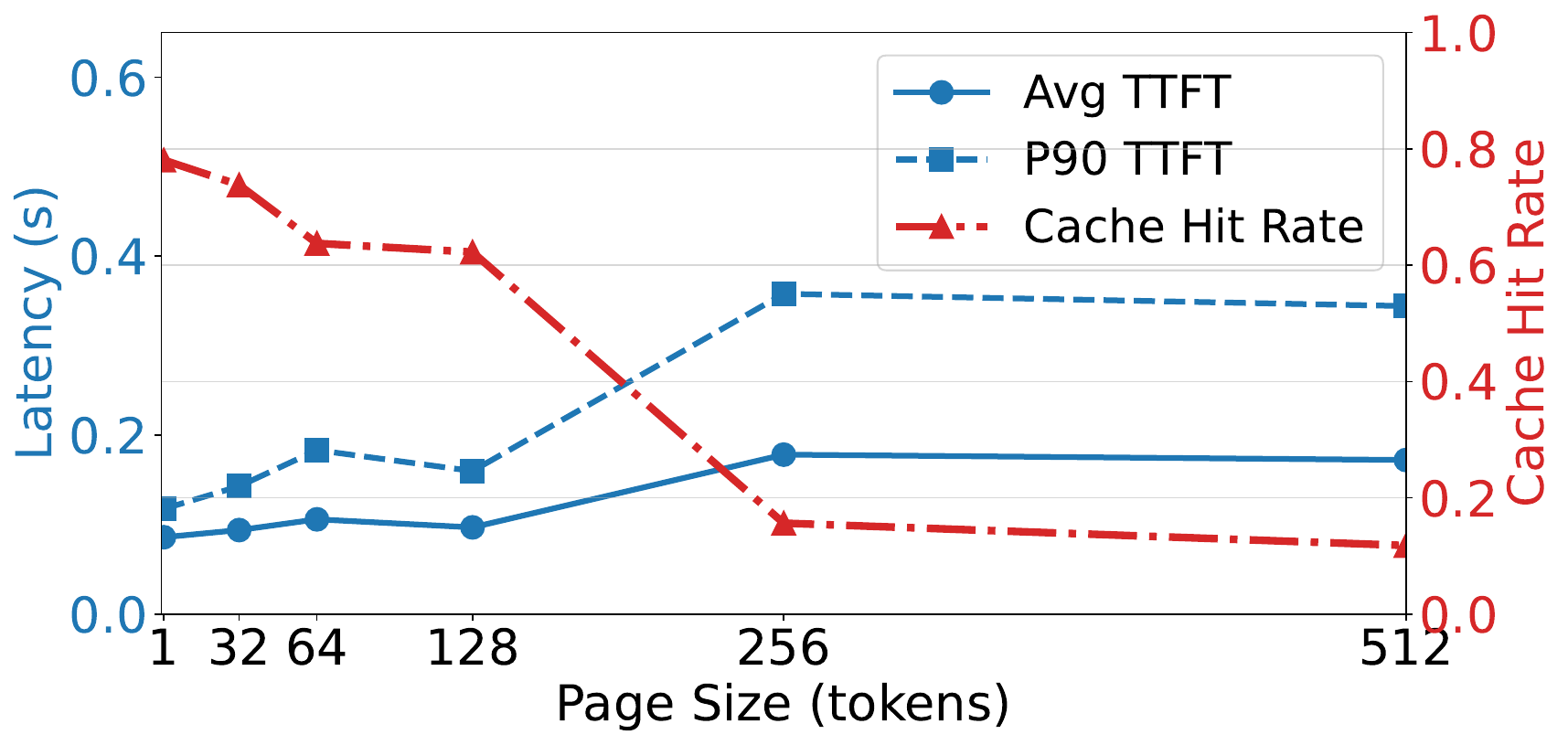}
    \caption{Large page sizes decrease cache hit rate and increase TTFT, benchmarked on H200 for Mistral-24B using the ShareGPT dataset.}
    \label{fig:page}
\end{figure}

\begin{figure}[t]
    \centering
    \includegraphics[width=.95\linewidth]{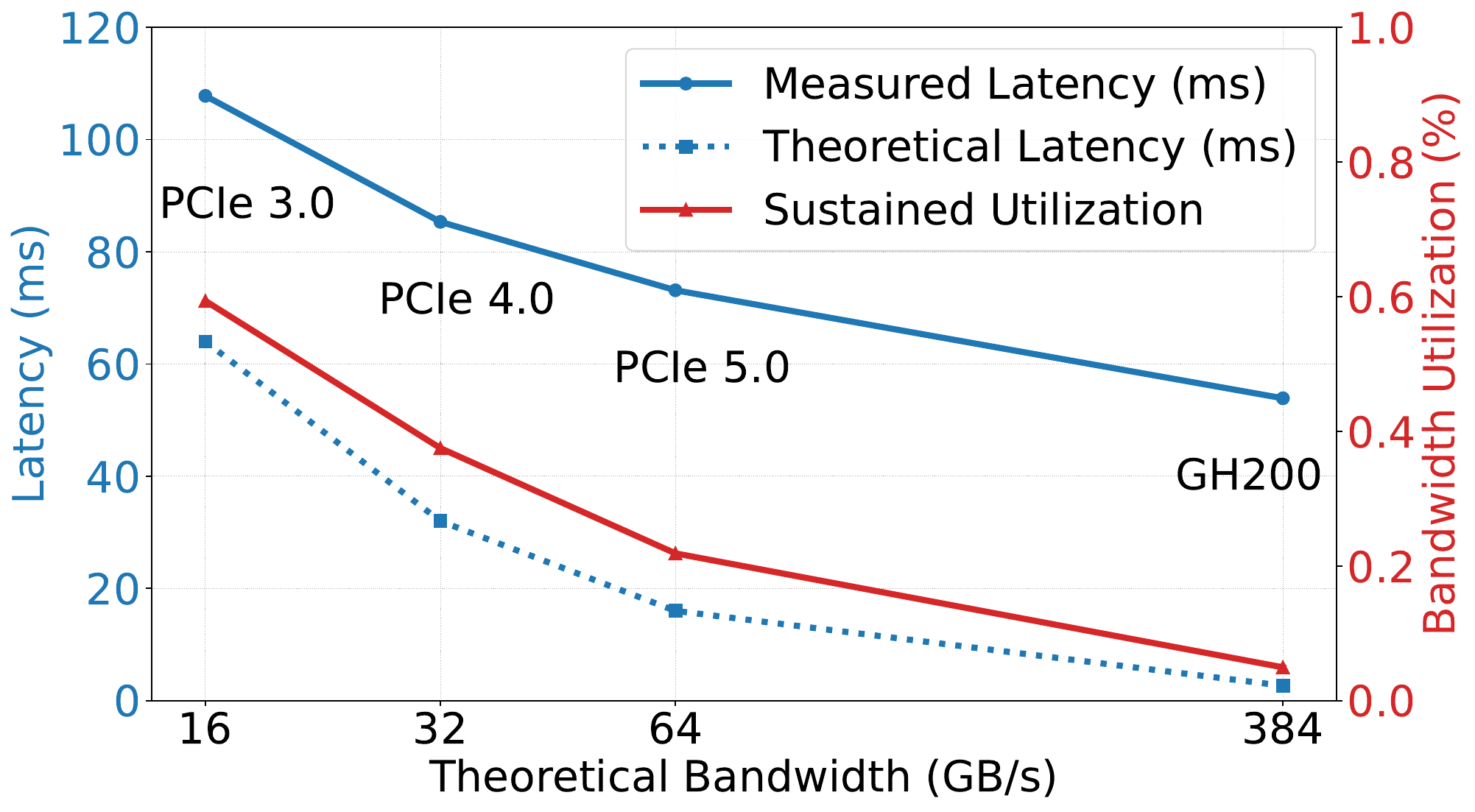}
    \caption{Latency and bandwidth utilization of loading KV caches of 8192 tokens (using page size 32) of Llama-3.1-8B from CPU to GPU on different platforms.
    }
    \label{fig:loading}
\end{figure}

\subsection{Low Bandwidth Utilization in KV Cache Transfers}\label{motivation:bandwidth}
The achievable throughput of an I/O subsystem is fundamentally constrained by the relationship described by Little's Law.
Let $\lambda$ be the arrival rate of I/O operations, $C$ be the average number of concurrent I/O operations, and $L$ be the average latency per operation.
According to Little's Law, we have $C = \lambda \cdot L$.
Let $X$ represent the sustained data throughput (e.g., in GB/s), $S$ be the average data size per I/O operation, the throughput can be described as $X = \lambda \cdot S$. Combining with Little's law we have $X = C\cdot S/L$ describing attainable throughput in a stable state.
This equation highlights that maximizing data throughput ($X$) requires either high concurrency ($C$), large transfer sizes ($S$), or low latency ($L$).

To transfer data between CPU and GPU memory, asynchronous operations like cudaMemcpyAsync are typically used to engage the GPU's Direct Memory Access (DMA) engine~\cite{emogi}. 
However, the latency ($L$) in this scenario includes non-negligible CPU-GPU communication overhead and scheduling delays~\cite{hwang2023ark}. While concurrency ($C$) can be increased by launching numerous asynchronous I/O operations, practical limits arise from the available application-level parallelism (e.g., on the CPU) and the queue capacities within the GPU driver (e.g., CUDA driver) and hardware~\cite{emogi,bam}.
Consequently, increasing the transfer size ($S$) often becomes the most practical lever for improving bandwidth utilization.
Saturating modern high-bandwidth interconnects underscores this point; for example, achieving 75-80\% of theoretical PCIe 5.0 bandwidth necessitates transfer sizes ($S$) in the megabyte range (i.e., 1-2MB).
This principle is not limited to CPU-GPU transfers; achieving high throughput on other media like SSDs or network interfaces often demands even large transfer sizes~\cite{eisenman2019flashield, kangaroo, qizhensdi}.

\noindent\textbf{Cost of Large Pages.} 
However, contrary to requirement of transfer efficiency, LLM inference systems favor small granularity.
Smaller pages, e.g., in the range of 1-32 tokens, generally lead to better memory utilization~\cite{sosp23:kwon_vllm} and can improve cache hit rate~\cite{nvidia2024tensorrt-blog}, as cache matching is performed on a per-page basis.
While increasing the page size (i.e., compacting more tokens into a single continuous page and effectively increasing $S$) may appear to improve bandwidth utilization based on the formula, it introduces substantial performance penalties for LLM inference. 
To demonstrate this trade-off, we benchmarked a Mistral-24B model using the popular ShareGPT dataset, varying the KV cache page size from 1 to 1024 on the SGLang framework. 
As shown in Figure~\ref{fig:page}, increasing the page size leads to a significant drop in the KV cache hit rate. This degradation directly results in a substantial increase in both the average and P90 TTFT, which rise by up to $2\times$ and $2.9\times$, respectively, at the largest page sizes tested.
We also observed similar trends in long context benchmarking scenario as shown in Figure~\ref{fig:loogle_page}.

This preference for small pages results in significantly small effective transfer sizes ($S$) for KV cache operations, 
leading to severely underutilized I/O bandwidth. 
As illustrated in Figure~\ref{fig:loading}, transferring KV cache data for 8192 tokens, achieves only approximately 22\% of the theoretical PCIe 5.0 bandwidth. 
Page size is set to 32, a value recommended in prior works on hierarchical KV cache~\cite{arxiv24:hu_memserve, atc24:gao_cachedattention, asplos25:flashgen, eurosys25:yu_pensieve} and a maximum supported size in vLLM for CUDA GPUs~\cite{vllm_cache_config}.
This underutilization is exacerbated on platforms with even higher interconnect bandwidths, falling to as low as ~5\% on systems like NVIDIA's Grace-Hopper platform that replaces PCIe with NVLink and offers 6x higher peak bandwidth.
This fundamental trade-off between transfer efficiency and caching benefit underscores the need for a more effective I/O mechanism that can achieve the best of both worlds.

\subsection{Resource Orchestration and Delay Hits}
\label{motivation:schedule}
Typical LLM serving engines often treat GPU compute and HBM as first-class resources, since sub-tasks are usually either compute-bound (e.g., dense and attention computation in prefill) or memory-bound (e.g., decoding). 
With the introduction of hierarchical KV caching, this picture largely remains unchanged. Existing works either regard KV cache loading from CPU to GPU as negligible~\cite{eurosys25:yu_pensieve, atc24:gao_cachedattention}, assuming the latency of loading cache data of layer $N+1$ via PCIe can be effectively hidden by overlapping I/O with the computation of layer $N$; or else opt for recomputation when the cost is justified~\cite{jin2024compute}.
However, when serving requests with long cached contexts but relatively few new tokens to prefill, this assumption breaks down. 
On one hand, the latency of bulk KV cache transfers can exceed what layer-level computation can hide; 
on the other hand, recomputation becomes increasingly costly as context length grows, making it an unattractive alternative. 
Consequently, layer-wise overlapped prefill can degrade into a PCIe bandwidth–bound task.
Figure~\ref{fig:stall} illustrates this bottleneck: even with an I/O mechanism achieving 75\% of theoretical PCIe bandwidth, stalls still account for up to 24\% of prefill execution time. 

Compounding this challenge is the \emph{delay hit phenomenon}~\cite{delayhit}, previously studied in the networking community, which arises when multiple requests for the same data object arrive and queue while an initial cache miss is still being resolved. 
We observe an analogous effect in LLM serving: under high traffic, multiple requests may target the same (or a prefix of the same) context. 
As illustrated in Figure~\ref{fig:scheduling}, when such requests are grouped into the same batch, redundant prefill computation occurs. This problem is exacerbated by the asynchronous schedulers widely adopted in modern serving engines~\cite{sglang_scheduler,zhu2025nanoflow}, which prepare the next batch before the ongoing one completes, extending the cache-miss resolution window to the full execution time of a batch and making delay hits more likely. 
The impact is especially severe in long-context scenarios, where both recomputation cost and cache-miss resolution time scale unfavorably. 

These observation highlight the need to re-think scheduling policies in LLM serving. In particular, schedulers must explicitly treat CPU-GPU bandwidth as a first-class resource, balancing computation and data transfer when batching requests, more strategically overlapping I/O with compute, and mitigating delay hits to avoid redundant work while sustaining throughput.
\section{\oursys's Design and Implementation}

\begin{figure}[t]
    \centering
\includegraphics[width=\linewidth]{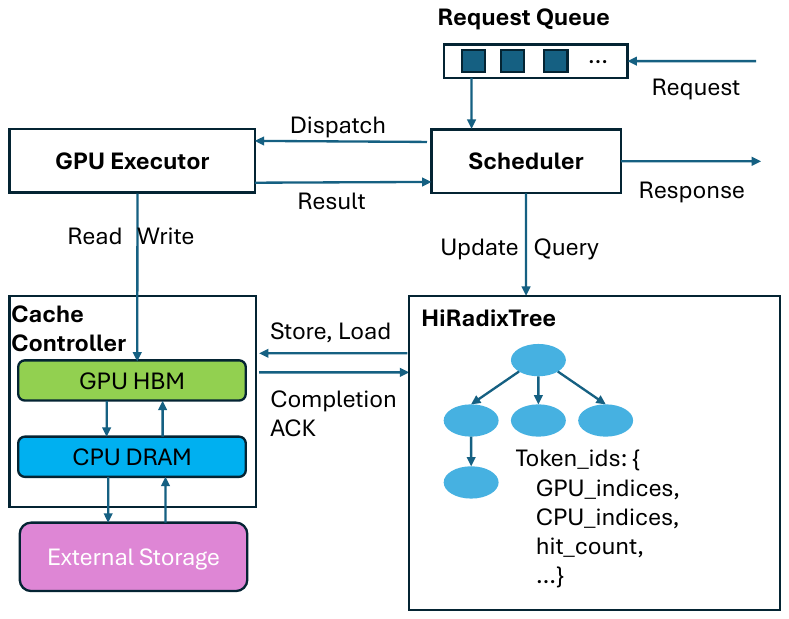}
    \caption{System Architecture of \oursys}
    \label{fig:system}
\end{figure}

\subsection{Overview}\label{sec:design_overview}
Motivated by challenges discussed in \S\ref{sec:motivation}, we built \oursys, a system with two key components.
The \oursys \textit{Cache Controller} (\S\ref{sec:design_io}) manages the data plane elements throughout the memory hierarchy.
It introduces an optimized GPU-CPU data transfer mechanism and manages KV cache memory layouts to support efficient small page transfers as motivated in \S\ref{motivation:bandwidth}.
The \oursys \textit{Scheduler} (\S\ref{sec:design_scheduling}) implements the control plane that intelligently schedules requests in a cache resource-aware manner as motivated in \S\ref{motivation:schedule}.
It references a \textit{HiRadixTree}, which is an extension to SGLang's RadixTree~\cite{neurips24:zheng_sglang}, effectively serving as a page table and stores metadata about each KV cache page.

Figure~\ref{fig:system} presents the \oursys architecture.
When a request is submitted, it enters a request waiting queue. 
During the execution of the ongoing batch, the \textit{Scheduler} continuously estimates available system resources and the resource demands of queued requests, and selects a subset to form the next batch.
The \textit{Scheduler} then sends this batch to GPU executor and initiates a KV cache loading request to the \textit{Cache Controller}.
During the execution of the prefill batch, the GPU executor synchronizes with \textit{Cache Controller} to ensure that the KV cache of certain layer is available before the execution.
Once prefill is complete, the prefilled requests are merged into a consolidated decoding batch via continuous batching~\cite{orca}. 
\oursys uses a P-D co-location design, alternating the execution of prefill and decoding batches temporally on the same GPU, and follows SGLang's practice to prioritize the execution of prefill batch for shorter response time (TTFT) and to form a larger decoding batch for higher throughput.
Finally, the \textit{Cache Controller} actively manages the backup and eviction of any KV cache pages to lower memory hierarchies asynchronously.

\subsection{Efficient KV Cache I/O}\label{sec:design_io}

To address the limitations discussed in \S\ref{motivation:bandwidth}, inspired by established practices within the computer architecture community~\cite{emogi, bam}, \oursys leverages \textit{GPU-assisted I/O} to transfer KV cache pages between CPU and GPU memory for low-latency I/O on small, fragmented data.
Specifically, instead of invoking standard \emph{cudaMemcpyAsync} API repetitively with small data transfers, a GPU-assisted I/O job operates by launching a CUDA kernel. This kernel spawns thousands of threads. Each thread is responsible for loading a small chunk of data from a source (either GPU global memory or CPU registered pinned memory) into its local register files and then streaming this data to a destination (which can also be GPU global memory or registered CPU pinned memory).

GPU-assisted I/O offers several advantages:
First, it enables \textbf{enhanced concurrency} ($C$): GPUs provide massive, cost-effective parallelism, supporting thousands of concurrent I/O operations compared to typically only tens on CPUs.
Second, it is \textbf{compatible with small transfers} (efficient $S$): the granularity required for efficient GPU-assisted I/O is only 128 bytes on most architectures~\cite{nvidiaPTX}, which is sufficiently fine for single-page KV caches (kilobytes), eliminating the need to inflate page size for efficiency.
Finally, it allows \textbf{flexible memory layout}: since light computation in I/O kernels is virtually free, layout transformations between GPU and CPU memory can be performed at negligible cost, enabling flexible and efficient data organization (see \S\ref{design:layout} for details).

\begin{figure}[t]
    \centering
    \includegraphics[width=\linewidth]{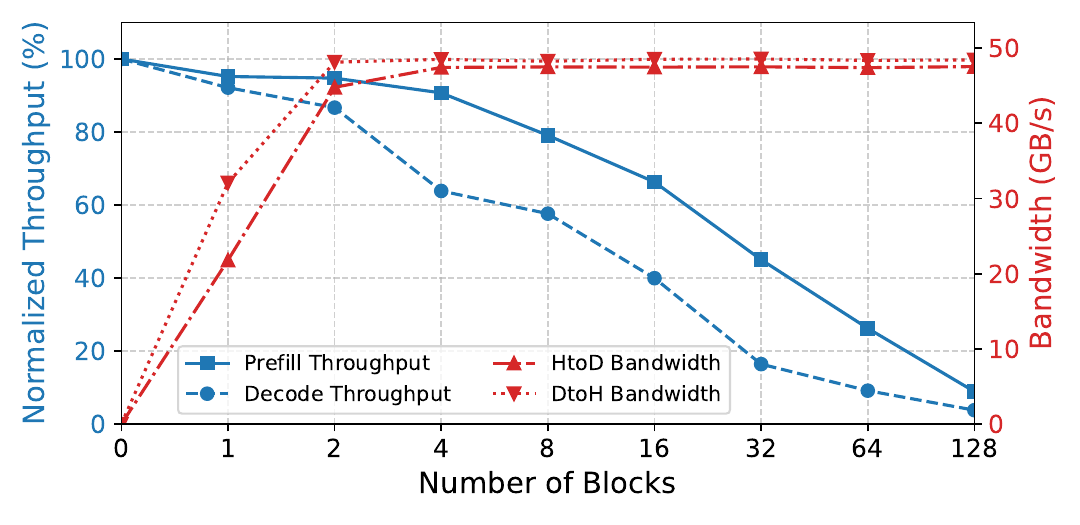}
    \caption{
    Performance interference vs. resources allocated to the KV-cache I/O kernel. Measurement on concurrently running \oursys’s I/O kernel with a prefill pass (batch of two requests with 4k input each) and a decode pass (batch of 16 requests with 4k input each), respectively.
    }
    \label{fig:interference}
\end{figure}

However, a challenge associated with GPU-assisted I/O, as highlighted in prior work~\cite{hwang2023ark}, is runtime interference when co-running with other kernels. 
Without dedicated hardware handling the fine-grained I/O tasks, GPU threads consume valuable resources, such as register files and execution cycles, and can lead to cache pollution.
Prior work~\cite{rammer} also demonstrated that GPU hardware schedulers often struggle to effectively manage this resource contention, potentially degrading the performance of both the I/O operations and concurrent computational kernels.

We observe that efficient data transfer does not need to monopolize the entire GPU.
\oursys employs a strategy of launching a small number of large CUDA blocks to incentivize the GPU's hardware scheduler to confine these I/O kernels to a small subset of Streaming Multiprocessors (SMs), as few as 1.
This targeted allocation, when combined with low-level instructions to bypass the cache and thereby mitigate pollution, minimizes interference with concurrent workloads.
Moreover, with the ROCm backend~\cite{amd_yes}, these kernel implementations are also compatible with AMD GPUs.
To balance resources for overall efficiency, we conducted microbenchmarks co-running I/O kernels with prefill and decoding kernels on an NVIDIA H200 GPU. 
As shown in Figure~\ref{fig:interference}, using only two CUDA blocks of 1024 threads each, \oursys achieves nearly 50~GB/s transfer throughput while incurring less than 5\% performance degradation on prefill and 10\% on decoding. Based on these results, we select two blocks as the default quota for loading data from CPU to GPU (a critical path operation), and one block for backing up data from GPU to CPU (a non-critical path), 
where the bandwidth is already sufficient and overhead must be minimized.
Our end-to-end evaluation confirms that this configuration sustains high I/O bandwidth while keeping overall performance impact under 5\%, demonstrating that carefully tuned GPU-assisted I/O can be both efficient and non-intrusive.

\begin{figure}[t]
    \centering
\includegraphics[width=\linewidth]{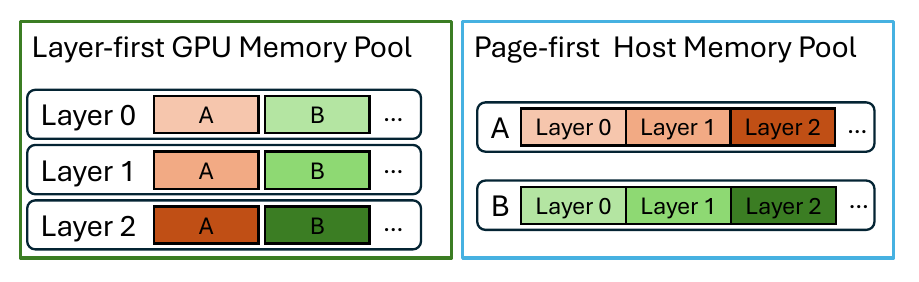}
    \caption{Layer-first v.s. Page-first layouts}
    \label{fig:layout}
\end{figure}

\subsubsection{Data Management Beyond Host Memory}
\label{design:layout}
When external storage is involved, the cache controller opportunistically prefetches data from storage to host memory when a cache hit is detected at the storage layer. 
The latency of this prefetch overlaps the request’s queuing delay.
Once the scheduler dispatches the request for execution, the cache controller terminates any in-flight prefetch and leverages the available cache already in host or GPU memory. 
This best-effort approach is motivated by the significantly higher and less predictable latency of the storage layer compared to the data transfers between host and GPU memory, which \oursys adopts a layer-wise overlapping approach.

Furthermore, the data transfer inefficiency caused by fragmented memory layout, as motivated in \S\ref{motivation:bandwidth}, also extends to other storage media. 
In addition to small pages, LLM serving systems also favor a layer-first memory layout in the GPU memory pool (shown in Figure~\ref{fig:layout}), as it aligns with the layer-wise nature of LLM computation.
However, this layout further fragments data, hindering bulk data transfer efficiency. 
An alternative, transfer-friendly layout that arranges layers of a page contiguously would be ideal for I/O but would require an layer of indirection for computation, complicating kernel implementation.

By leveraging GPU-assisted I/O, \oursys resolves this conflict by enabling a virtually free, on-the-fly transformation between the compute-friendly and transfer-friendly layouts. 
To perform the layout transformation, a thread simply applies one additional arithmetic operation to its assigned offset to calculate the correct destination address.
This operation has negligible overhead. As illustrated in Figure~\ref{fig:layout}, this capability decouples the layout requirements across the memory hierarchy: the GPU can maintain its computation-friendly layer-first layout, while other media, such as host memory and external storage, can adopt a page-first layout that maximizes transfer efficiency with larger, contiguous data blocks. 
In \S\ref{eval:layout}, we demonstrate how this decoupled layout strategy significantly reduces data loading time.

\begin{figure}[t]
    \centering
\includegraphics[width=\linewidth]{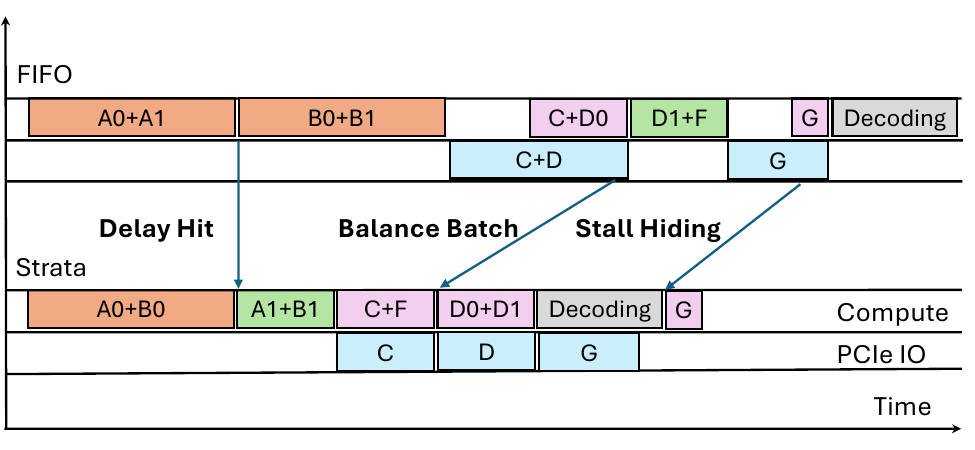}
    \caption{Scheduling Policies of \oursys, where orange blocks indicate prefill batches experiencing cache miss, green indicates cache hit on device, purple indicates cache hit on host memory, blue indicates data transfer, and the one decoding batch is colored in gray.
    }
    \label{fig:scheduling}
\end{figure}

\subsection{Cache Aware Scheduling}\label{sec:design_scheduling}

As motivated in \S\ref{motivation:schedule}, the goal of the scheduler is to maximize caching benefit by avoiding delay hit and loading stall.
The \textit{Scheduler} does so through three stages.
First, it identifies requests that are potentially susceptible to delay hits and \textbf{defer the execution} to right after the delay hit resolved.
This eliminates unnecessary cache miss without impacting on TTFT.
Secondly, it \textbf{formulates a balanced batch} that aims to pair with loading (from host memory) with sufficient computation to hide the loading latency.
Finally, in the event that batches are still loading-bound, the \textit{Scheduler} \textbf{hides I/O stalls} by inserting useful compute inside bubbles.

\subsubsection{Deferral on Delay Hit}
\label{schedule:delay}
As discussed in \S\ref{motivation:schedule}, delay hits can cause redundant computation in two scenarios: (i) when multiple requests sharing the same cache miss are scheduled into the same batch (Figure~\ref{fig:scheduling}), and (ii) when the execution of a request is prepared asynchronously without awareness that the corresponding context cache is still being computed.
To keep track of potential delay hit, we introduce \emph{transient nodes} in the HiRadixTree. Similar to standard nodes, they use token IDs as traversal keys, but instead of pointing to memory indices, they carry one of two marks: \texttt{in-queue}, indicating that a request is referencing a new context, and \texttt{in-flight}, indicating that the cache for the corresponding tokens is under computation. When iterating over the request queue, \oursys inserts transient nodes marked \texttt{in-queue} as needed. If a request matches existing transient nodes, it is deferred to the next scheduling round but placed at the front of the waiting queue to benefit from the soon-to-be-hot cache and minimize its impact on TTFT.

When a request proceeds to execution, its associated transient nodes are marked \texttt{in-flight}. Upon completion, the nodes are converted into standard nodes, with indices pointing to the ready context cached in memory. 
To prevent unnecessary deferrals, \oursys uses a configurable threshold: a request is deferred only when the number of token matches on transient nodes exceeds this value. In practice, a default threshold of 100 active token matches proved effective.

\begin{algorithm}[t]
\caption{Balanced Batch Formation}
\label{alg:scheduler}
\begin{algorithmic}[1]
\Procedure{AddBundleHit}{$Q,B$}
  \For{each $r$ in $Q$}
    \If{$B.\mathrm{is\_bundle\_hit}(r)$}
      \State $B.\mathrm{add}(r)$; $Q \gets Q - r$
    \EndIf
  \EndFor
\EndProcedure
\Function{BatchFormation}{$Q$}
  \State $B\gets\mathrm{Batch}()$; $D\gets[]$
  \State $B.\mathrm{add}(Q.\mathrm{pop}(0))$; \Call{AddBundleHit}{$Q,B$}
  \While{$|Q|>0$ \textbf{and} $\neg B.\mathrm{is\_full}()$}
    \State $r\gets Q.pop(0)$
    \If{$B.\mathrm{loading\_bound}(r)$}
      \State $D.\mathrm{append}(r)$
    \Else
      \State $B.\mathrm{add}(r)$; \Call{AddBundleHit}{$Q,B$}
    \EndIf
  \EndWhile
  \For{each $r$ in $D$}
    \If{$B.\mathrm{is\_full}()$} \textbf{break} \EndIf
    \State $B.\mathrm{add}(r)$
  \EndFor
  \State \Return $B$
\EndFunction
\end{algorithmic}
\end{algorithm}

\subsubsection{Balanced Batch Formation.}
After removing candidates susceptible to delay hits, the scheduler selects requests to form the next prefill batch.
In most LLM serving engines~\cite{neurips24:zheng_sglang, sosp23:kwon_vllm}, batch formation follows a FIFO policy by default, where requests are taken in arrival order until the batch is full (either reaching a preset token limit or exhausting GPU memory).
To address the loading-bound issue discussed in \S\ref{motivation:schedule}, \oursys introduces a new batch formation mechanism that balances data loading with sufficient computation. 
An example is illustrated in Figure~\ref{fig:scheduling}, a batch containing requests C and D0 would require loading both contexts, causing a loading stall. 
In contrast, forming a different batch (C, F) could get the loading of C overlapped.
A special case worth noting is the other new batch (D0, D1): since they share the same context, batching them not only balances loading with compute but also reduces GPU memory usage and on-device bandwidth pressure, further improving efficiency. 
We refer to this as a \textit{bundle hit}, on the opposite of \textit{delay hit}.

The procedure is detailed in Algorithm~\ref{alg:scheduler}. Before each batch is formed, the scheduler obtains the \textit{load} and \textit{compute} requirements of each request using the HiRadixTree. 
During batch formation, as it iterates through the queue, the scheduler checks whether adding a request would reach the loading-bound limit (`loading\_bound' in line 11), defined as the ratio of aggregated load to compute. 
When this ratio exceeds a threshold, the batch is considered loading-bound.
This threshold is hardware- and model-dependent and thus can be profiled separately; in practice, \oursys uses a default ratio of 100, corresponding to the point where stalls begin to appear showed in Figure~\ref{fig:stall}.
If the request would fit into the batch without making it loading-bound, it will be added into the batch, then the scheduler will iterate through all the rest requests to preferentially add requests that bundle-hit with it. 
Otherwise, the request is moved to a deprioritized list. 
If the batch is not full till the end of the queue, the scheduler supplements it with deprioritized requests (line 17).
To prevent starvation, deprioritized requests retain their original order, and each batch formation always begins with the first request in the queue.

\subsubsection{Hide Loading Stall with Bubble Filling}
Even with balanced batching, some batches can still be loading-bound. 
The final strategy of the scheduler is \textit{bubble filling} that overlaps loading stalls with useful computation.
An example is illustrated in Figure~\ref{fig:scheduling}, when request G requires a long context load, the scheduler defers computation of the prepared prefill batch and instead issues a decoding batch to the model executor to run concurrently with the context loading.
This strategy complements SGLang’s default prefill-first policy (as discussed in \S\ref{sec:design_overview}), allowing some flexibilities when choosing between prefill and decoding for improved overall utilization.
Although decoding batches are also I/O-bound, they primarily saturate HBM bandwidth, whereas loading tasks saturate PCIe bandwidth. This distinction enables the two operations to overlap with minimal resource contention.
It is also possible to insert an prefill batch to fill the bubble if available, which will be more applicable to P-D disaggregated systems~\cite{zhong2024distserve}.
\section{Evaluation}
\subsection{Methodology}
\noindent\textbf{Testbed.}
We evaluate \oursys and baselines on two platforms. 
The \textit{H200} platform is a node equipped with 8 NVIDIA H200 GPUs interconnected with NVLink, an Intel Sapphire Rapids CPU, and 1.6TB of DRAM.
Each GPU is connected to the CPU via a PCIe 5.0 x16 link, offering up to 64 GB/s of peak bandwidth (unidirectional).
The \textit{GH200} platform is a GH200 Grace Hopper superchip~\cite{website:grace-hopper} node, which contains one NVIDIA H100 GPU integrated with one NVIDIA Grace 64-core ARM CPU.
The \textit{GH200} system is equipped with 464GB of LPDDR5X DRAM, providing up to 384 GB/s of memory bandwidth (unidirectional) to the CPU.

\noindent\textbf{Baselines.}
We compare \oursys with following state-of-the-art baselines.

\textit{vLLM}~\cite{sosp23:kwon_vllm} is a popular open-source serving engine. 
Additionally, \textit{vLLM-LMCache} enables hierarchical caching on vLLM using the official community extension of LMCache~\cite{website:lmcache}.
For our benchmarks, we used vLLM v0.8.5 and LMCache v0.2.1.
The LMCache chunk size is set to 256 as default and vLLM page size was set to 32 in line with prior work~\cite{atc24:gao_cachedattention}.

\textit{TensorRT-LLM}~\cite{website:tensorrt} is an open-source serving library from NVIDIA, specialized for NVIDIA GPUs.
Additionally, \textit{TensorRT-HiCache} enables hierarchical caching on top of TensorRT-LLM through its automatic CPU memory offloading feature.
We used TensorRT-LLM v0.17.0 in our benchmarks, with the page size also set to 32 as default.

\textit{SGLang}~\cite{neurips24:zheng_sglang} is an open-source serving engine that delivers comparable performance to vLLM, while offering a more lightweight and customizable architecture. 
To enable a direct comparison to \oursys, we implemented \textit{SGLang-HiCache} which incorporates a state-of-the-art layer-wise KV cache transfer overlapping and hierarchical caching implementation using \emph{cudaMemcpyAsync} transfers, which is in line with prior work including CachedAttention~\cite{atc24:gao_cachedattention}, Pensieve~\cite{eurosys25:yu_pensieve} and FlashGen~\cite{asplos25:flashgen}.
We used SGLang v0.4.5 for all three systems.
We set the page size for \textit{SGLang} and \oursys to 1 (SGLang's default), and the page size for \textit{SGLang-HiCache} to 32 to be consistent with other hierarchical cache baselines.

\noindent\textbf{Models.}
We utilize three popular open-source LLMs with long context capabilities, spanning small, medium, and large sizes: Llama-3.1-8B-Instruct~\cite{llama3.1} (128k context window), Qwen2.5-14B-Instruct-1M~\cite{qwen2.5-1m} (1M context window), and Llama-3.1-70B-Instruct~\cite{llama3.1} (128k context window).
We served the 8B and 14B models using a single GPU, and served the 70B model using 4 GPUs configured with tensor parallelism.
\begin{table}[t]\small
\centering
\begin{tabular}{lrrrr}
\hline
                    & LooGLE & NarrativeQA & ReviewMT & ShareGPT \\
\hline
avg. in   & 21613  & 54797   & 17708    & 680.9     \\
avg. out  & 15.60  & 13.00  & 208.3     & 260.9     \\
\# contexts         & 105    & 50          & 100        & - \\
\# queries          & 2410   & 1461       & 1092  & 200869   \\
\hline
\end{tabular}
\caption{Dataset statistics.}
\label{tab:dataset_comparison}
\vspace{-1mm}
\end{table}

\noindent\textbf{Datasets.}
We construct workloads from three long context datasets.
\textit{LooGLE}~\cite{loogle} features long documents from diverse sources such as arXiv, Wikipedia, and movie/TV scripts.
In our benchmarks, we use its Wikipedia portion, which provides both long and short queries paired with the documents.
\textit{NarrativeQA}~\cite{kovcisky2018narrativeqa}  is an influential long-context dataset for testing models' reading comprehension capabilities, featuring even longer context examples than LooGLE. We filtered documents exceeding 128k tokens because of context window limit of the test models, and sampled 50 documents from the remainder.
These two datasets mirror classic RAG use cases, in which extensive contexts are repeatedly queried by multiple users over time like question‑answering systems over technical manuals~\cite{gao2023retrieval}. 
\textit{ReviewMT}~\cite{tan2024peer} is a multi-agent conversation dataset, where agents simulate reviewers to converse about the quality of technical papers to make final decisions. This represents a typical agentic workflow involving long contexts.
We also include a dataset to evaluate \oursys's performance in short-context scenarios.
\textit{ShareGPT} is a popular conversational dataset comprised of a large collection of conversation histories from thousands of users, and was used in prior hierarchical KV caching studies~\cite{atc24:gao_cachedattention,eurosys25:yu_pensieve}.
Table~\ref{tab:dataset_comparison} summarizes the characteristics of these datasets.

Since individual query timestamps are not available in these datasets, we simulate query arrivals using a Poisson distribution to benchmark the system using varying request rates, following prior works~\cite{eurosys25:yu_pensieve,asplos25:flashgen}.
For conversational benchmarks (ReviewMT and ShareGPT), we preserve dependencies across conversation rounds. Consistent with the methodology in Pensieve~\cite{eurosys25:yu_pensieve}, we insert a 60-second “thinking time” between an LLM’s response and the user’s subsequent query for ShareGPT.
For the long-context benchmarks, queries are randomly sampled from the dataset. In \S\ref{eval:distance}, we further examine the performance characteristics under different workload patterns.
To avoid execution timeouts, we cap the maximum number of in-flight queries at 128 across all benchmarks. 
GPU memory is allocated according to each serving engine’s default policy to ensure fairness and performance. 
An exception is the ShareGPT dataset, where we restrict GPU memory to approximately 500K tokens to highlight the behavior of hierarchical caching baselines. 
For caching configurations that utilize CPU memory, we allocate 1 TB of system DRAM as pinned memory (400 GB on GH200 due to platform limits). 
Disk storage is not used in all benchmarks due to limited support in baseline systems.

\begin{figure*}[t]
  \centering
  \includegraphics[width=0.95\textwidth]{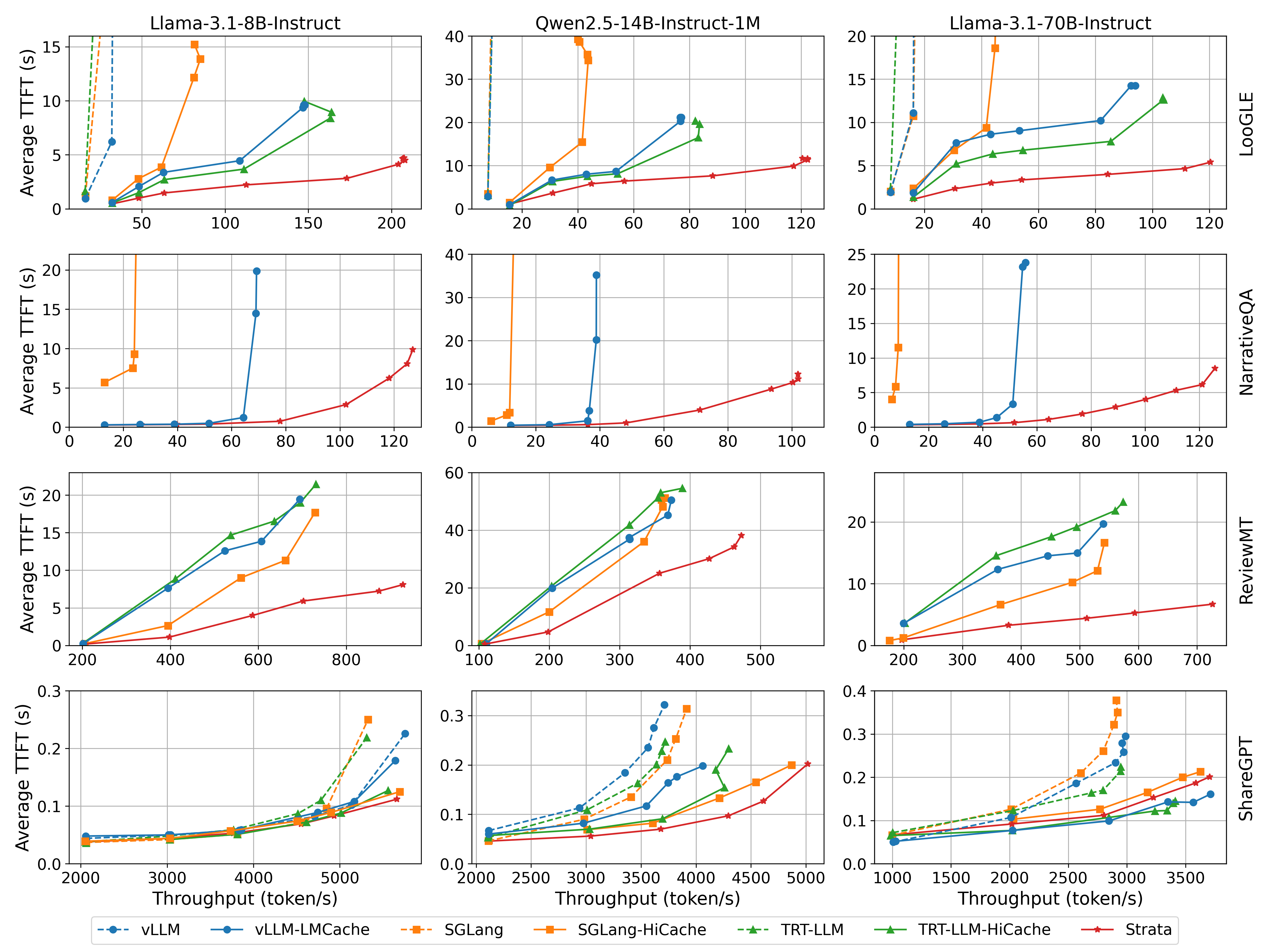} 
  \caption{
    End-to-end benchmark performance comparison on H200.  
  }
  \label{fig:e2e_all}
\end{figure*}

\subsection{End-to-end Performance Comparison}
\oursys is designed to improve long-context serving by reducing response latency and increasing overall throughput.
Accordingly, we evaluate the system using two primary metrics: average Time To First Token (TTFT) and output token throughput. TTFT captures query response time, a key determinant of user experience, while output token throughput is a widely adopted metric for characterizing LLM serving system performance~\cite{artificialanalysis_leaderboard}.

\subsubsection{How does the performance of \oursys compare to state-of-the-art LLM serving systems on long-context workloads?}
\label{eval:long}
Figure~\ref{fig:e2e_all} presents the achieved token throughput and corresponding average TTFT for each system under varying request rates across three models and four datasets.
First, we observe that \textit{hierarchical caching is essential for high-performance long-context serving}.
Across all three models, non-hierarchical caching solutions perform poorly on LooGLE. Without hierarchical caching, long-context workloads rapidly exhaust GPU memory capacity, leading to frequent cache misses during prefills. This, in turn, triggers repeated recomputation, resulting in low throughput and high latency. By contrast, systems equipped with hierarchical caching achieve substantially higher throughput and lower latency, consistently reaching approximately 95\% cache hit rate by leveraging CPU memory.

Second, \textit{\oursys delivers substantial improvements over existing hierarchical caching solutions}.
For Llama-8B on LooGLE, \oursys achieves up to $3.2\times$, $2.6\times$, and $1.9\times$ higher throughput at the same TTFT compared to \emph{SGLang-HiCache}, \emph{vLLM-LMCache}, and \emph{TensorRT-HiCache}, respectively. Similar gains are observed for Qwen-14B and Llama-70B: with Qwen-14B, \oursys achieves up to $3.9\times$, $2.1\times$, and $1.9\times$ improvements; with Llama-70B, the gains reach $5\times$, $5\times$, and $3.75\times$.
A consistent trend is seen on ReviewMT. Despite longer decoding reduces the dominance of prefill time, with Llama-8B, \oursys outperforms \emph{vLLM-LMCache} by $2.3\times$, \emph{TensorRT-HiCache} by $2.3\times$ and \emph{SGLang-HiCache} by $1.7\times$.
These performance gains stem from \oursys’s enhanced I/O efficiency and scheduling mechanisms, as further analyzed in \S\ref{eval:break}.

\subsubsection{How does \oursys perform with a warm cache at steady state?}
\label{eval:warmup}
NarrativeQA presents much longer average context lengths, resulting in an extensive prefill phase that made non-hierarchical caching solutions impractical.
Because this scenario presented the highest amount of memory pressure, we augmented this experiment to better understand steady-state performance characteristics (i.e., after the cache hierarchy has been filled)\footnote{NarrativeQA without pre-warming achieves similar results to the Loogle workload. We omit results for this scenario due to space.}.
In this setup, we first warmed up the CPU memory by pre-computing KV caches for all contexts in the evaluation set and then flush the KV cache on GPU memory to set initial state.
Second row of Figure~\ref{fig:e2e_all} reports the throughput-latency curve after restarting the workload post-warmup.
We report \oursys, \emph{SGLang-HiCache}, and \emph{vLLM-LMCache}, as \emph{TensorRT-HiCache} does not support pre-warming.
In this setting, \oursys achieves up to $2.3\times$, $2.6\times$ and $2.5\times$ throughput compared with \emph{vLLM-LMCache} on Llama-8B, Qwen-14 and Llama-70B models respectively.

\subsubsection{How does the performance of \oursys compare to state-of-the-art LLM serving systems on short-context workloads?}
\label{eval:short}
\oursys was explicitly designed for long-context workloads.
To understand the performance of \oursys on short-context workloads, the final row of Figure~\ref{fig:e2e_all} shows the average TTFT of the baseline systems, across three models, on the ShareGPT dataset.
Note that underlying SGLang engine exhibits a slight performance disadvantage compared to the base engines of vLLM and TensorRT-LLM on the Llama-8B and -70B models due to kernel differences.
Taking this into account, \textit{\oursys demonstrates comparable performance to the other state-of-the-art systems on short-context workloads}.

\subsection{In-depth Performance Analysis}
\label{eval:depth}
In this section, we conduct several breakdown analyses to better understand the performance benefits offered by \oursys. All analyses presented here were conducted using the Qwen-14B model on an H200 platform.

\begin{figure}[t]
    \centering
\includegraphics[width=\linewidth]{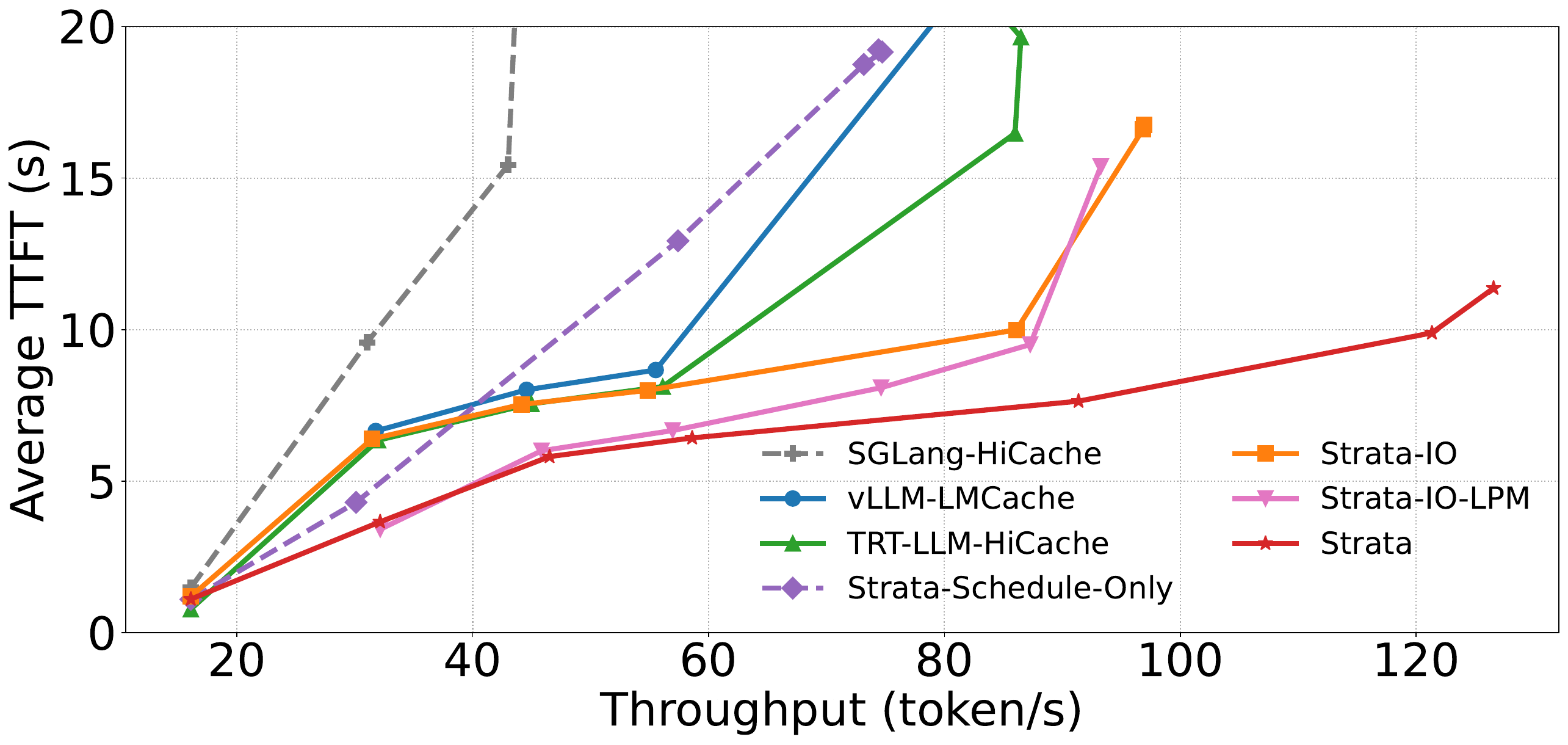}
    \caption{Breakdown of I/O and scheduling of \oursys.
    }
    \label{fig:breakdown}
\end{figure}

\subsubsection{How much does efficient I/O and scheduling benefit \oursys?}
\label{eval:break}
Figure~\ref{fig:breakdown} presents the throughput–latency curves of \oursys compared to three baselines.
On top of \textit{SGLang-HiCache}, we build and evaluate three ablated variants:
\textit{\oursys-IO}, which incorporates the GPU-assisted I/O mechanism from \S\ref{sec:design_io},
\textit{\oursys-Schedule-Only}, which applies the scheduling policy from \S\ref{sec:design_scheduling}, and \textit{\oursys-IO-LPM}, which integrates a longest prefix match (LPM) policy~\cite{neurips24:zheng_sglang}.

The results show that both the \emph{\oursys-scheduling} and \emph{\oursys-IO} components significantly improves the baseline hierarchical design, achieving up to $1.8\times$ and $2.3\times$ higher peak throughput, respectively. Each component alleviates the loading stall problem from a different perspective. Under low request rates, \emph{\oursys-scheduling} tends to deliver greater gains than \emph{\oursys-IO}, since smaller batch sizes generate lighter I/O pressure that can be more effectively mitigated by advanced scheduling. As the request rate increases, however, the I/O subsystem becomes the dominant bottleneck, making the GPU-assisted I/O mechanisms essential for sustaining high throughput.

We further compare \emph{vLLM-LMCache} and \emph{TensorRT-HiCache} directly with \emph{\oursys-IO}, since all three employ CUDA kernels to accelerate KV-cache I/O. As shown in Figure~\ref{fig:breakdown}, their performance is comparable at low request rates, but \emph{\oursys-IO} maintains higher throughput as the request rate rises, indicating more effective mitigation of interference at scale.
We also compare \oursys with \emph{\oursys-IO-LPM}, which increases the reuse count of on-device pages, thereby indirectly reducing host-side loading pressure and improving performance under low request rates. 
However, at higher request rates, it fails to sustain performance gains due to more frequent cache evictions. 
In contrast, \oursys consistently delivers improvements because it explicitly accounts for bandwidth resources in its design.

\begin{figure}[t]
    \centering
    \includegraphics[width=.9\linewidth]{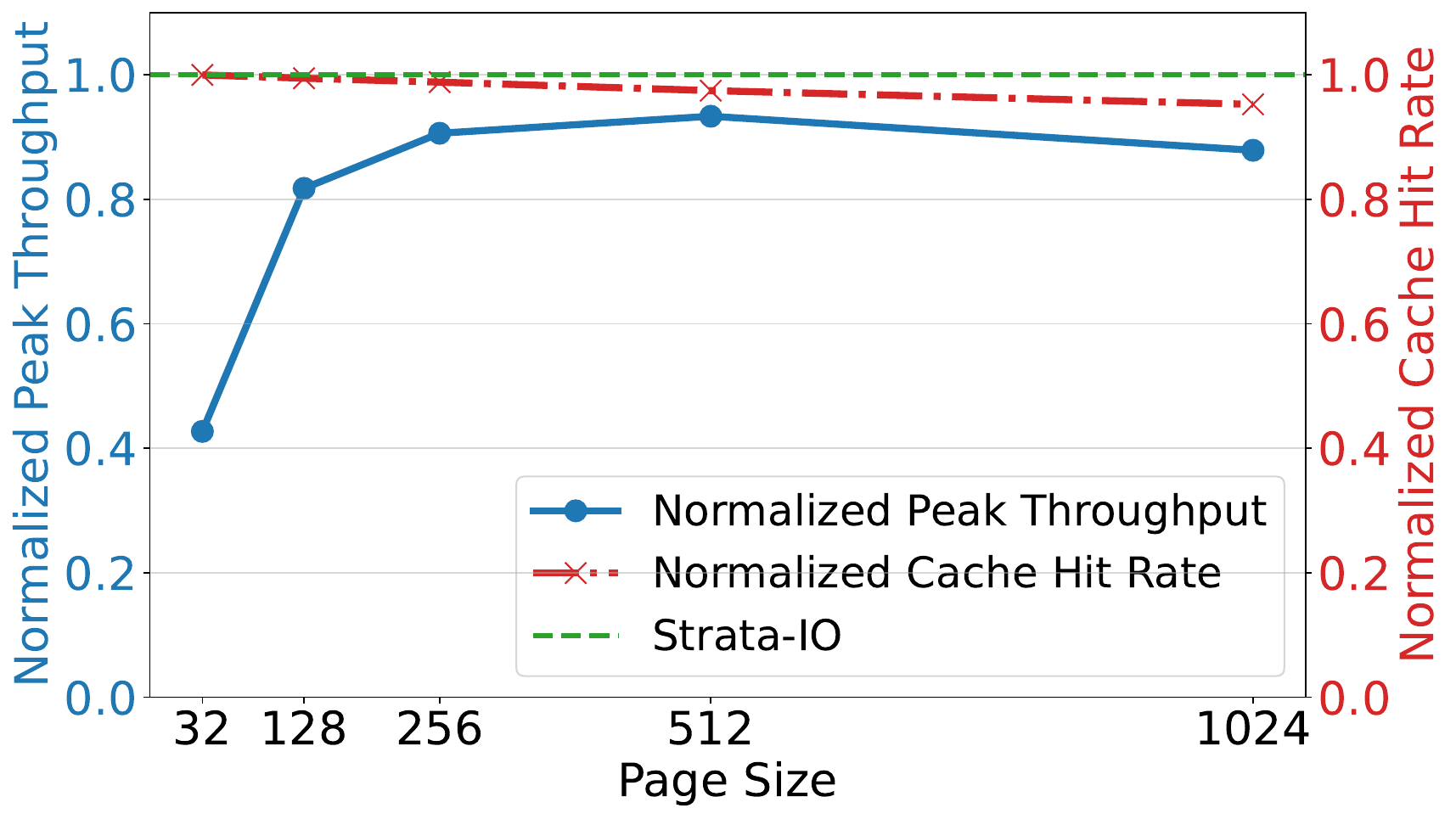}
    \caption{
    Performance comparison between \oursys+IO and SGLang+HiCache with different page sizes. 
    }
    \label{fig:loogle_page}
\end{figure}

\subsubsection{Can \oursys alleviate the burden of choosing a page size?}
\label{eval:page}
As discussed in \S\ref{motivation:bandwidth}, page size introduces inherent trade-offs among cache hit rate, I/O efficiency, and overall performance. This makes selecting the optimal page size a practical but non-trivial problem.
In contrast, with the GPU-assisted I/O mechanisms described in \S\ref{sec:design_io}, \oursys achieves consistently high I/O efficiency regardless of page size, effectively removing this burden from users.
To validate this claim, Figure~\ref{fig:loogle_page} reports the peak throughput of \textit{SGLang-HiCache} across different page sizes, normalized to \emph{\oursys-IO}. For \textit{SGLang-HiCache}, increasing the page size initially improves throughput, since larger pages reduce loading stalls. However, beyond a certain threshold, throughput declines as cache hit rate deteriorates. Even at its best-performing setting (page size 512), \textit{SGLang-HiCache} achieves only 93\% of \emph{\oursys-IO}’s performance, primarily due to a 2.4\% lower cache hit rate.

\subsubsection{Can \oursys adapt to varying cache distances?}
\label{eval:distance}
Cache distance is an important property of a trace for testing cache system performance.
To further augment the cache distance (i.e., access patterns between similar requests) of our workloads, we generate two additional workloads based on the LooGLE dataset.
In addition to our original (\textit{shuffle}) workload, we create a \textit{minimum cache distance} workload, where requests sharing the same context are lined up together, and a \textit{maximum cache distance} workload, where queries sharing the same context evenly distributed in the queue.

\begin{figure}[t]
    \centering
    \includegraphics[width=.9\linewidth]{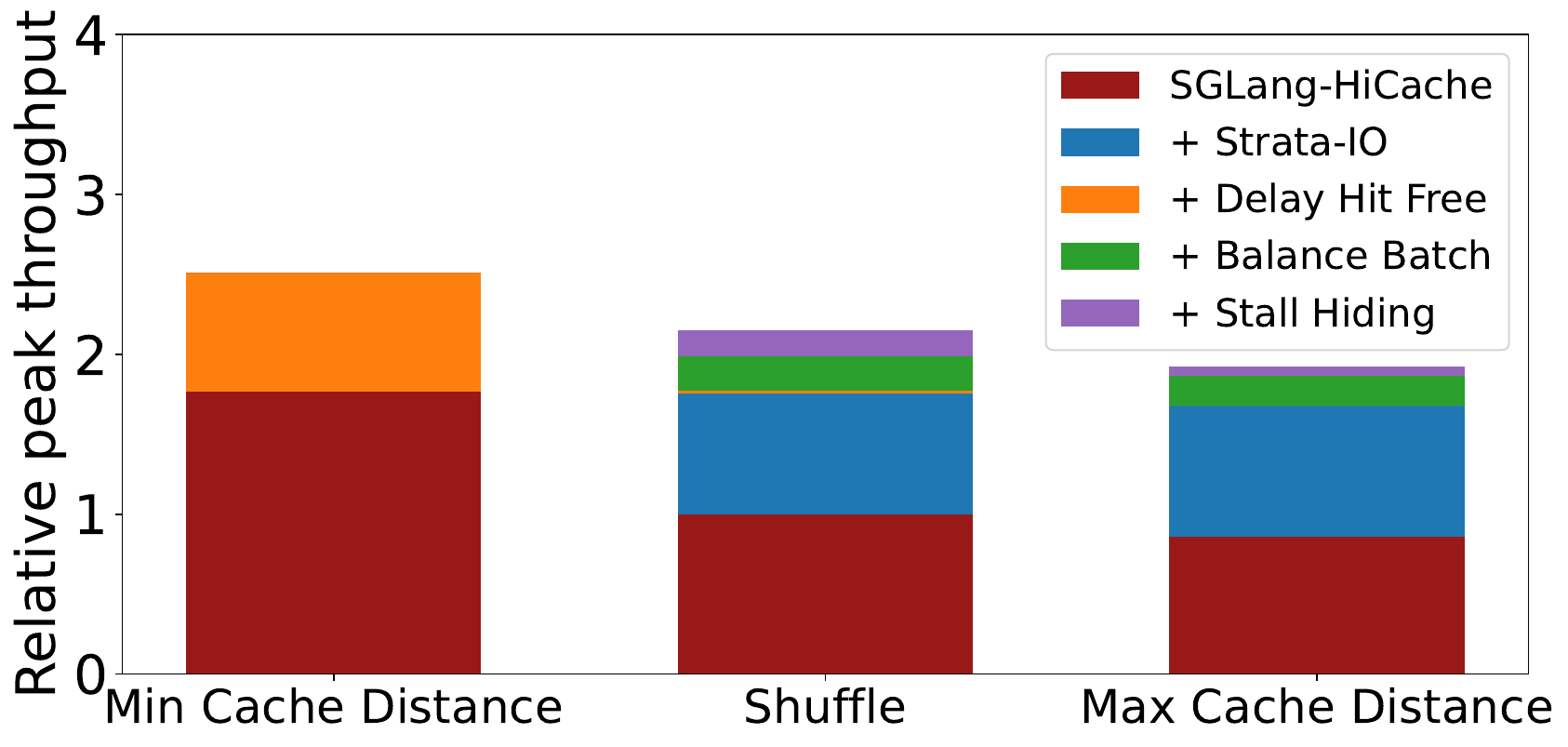}
    \caption{
    Breakdown of \oursys' optimization attributions on different workload patterns.
    }
    \label{fig:cache_distance}
\end{figure}

As shown in Figure~\ref{fig:cache_distance}, we observe that with minimal cache distance, there is no need for hierarchical caching due to the perfect locality.
In this scenario, delay hit mitigation benefits the most by improving effective cache hit rate, increasing peak throughput by $42\%$.
However, delay hit mitigation offers no benefit on max cache distance scenario because the high distance between similar requests naturally reduces the likelihood of the delay hit phenomenon.
In contrast,the I/O efficiency mechanisms presented in \S\ref{sec:design_io} result in an improvement of $76\%$ and $95\%$ for the shuffle and maximum cache distance, respectively, as larger cache distances result in more cache hits to CPU DRAM.
On top of that, balance batch further introduces $11\%$ and $12\%$ of the peak throughput improvement, respectively.
Stall hiding further introduce $8\%$ and $3\%$ for shuffle and max cache distance, with higher benefit to the shuffle pattern due to higher variance.

\begin{figure}[t]
    \centering
    \includegraphics[width=0.9\linewidth]{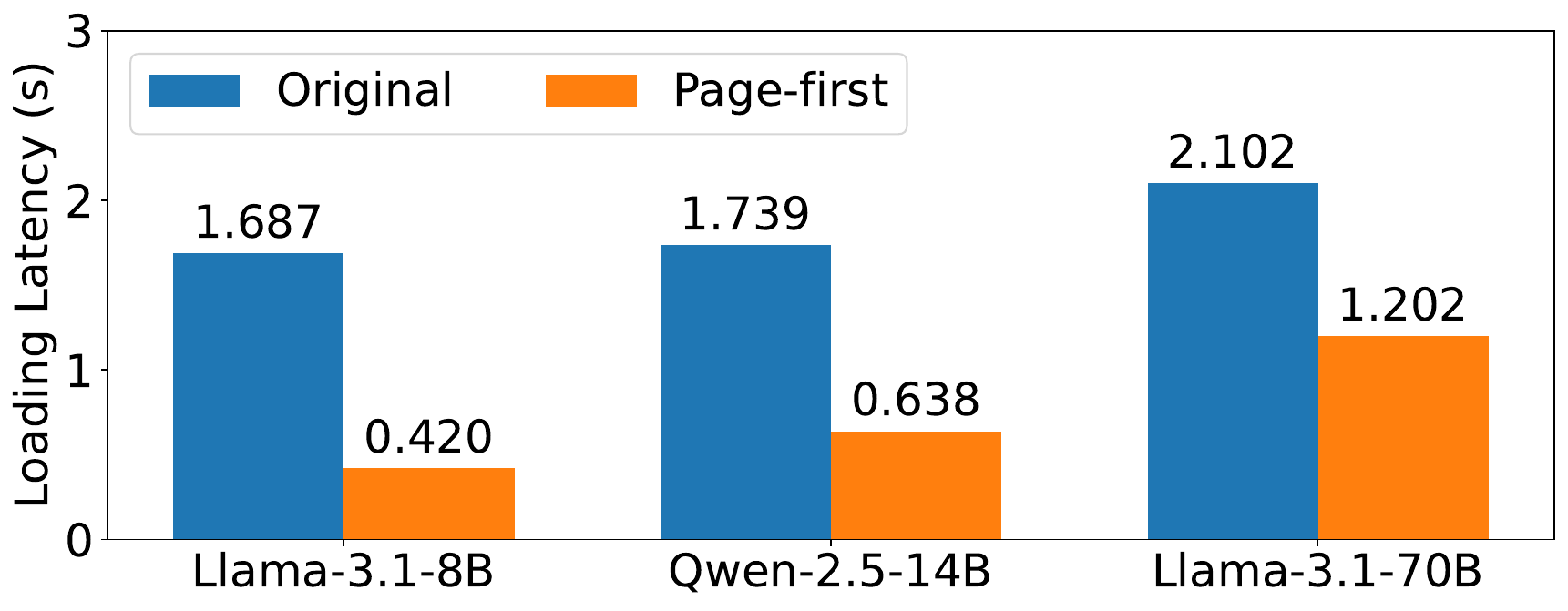}
    \caption{Latency of loading KV caches of 8192 tokens (using page size 32) of different models from a local disk to CPU memory with different memory layouts.
    }
    \label{fig:disk}
\end{figure}
\subsubsection{Does the decoupled memory layout benefit disk caching?}
\label{eval:layout}
GPU-assisted I/O enables the use of a page-first layout in CPU memory without requiring changes to the GPU memory layout, as discussed in \S\ref{design:layout}. 
Figure~\ref{fig:layout} presents a micro-benchmark demonstrating the benefit. 
Using the same page size of 32, the page-first layout achieves significantly larger transfer sizes, reducing latency by up to $4\times$ when loading 8192 tokens from disk.

\subsection{Benchmark on GH200 machine}
\begin{figure}[t]
    \centering
    \includegraphics[width=0.95\linewidth]{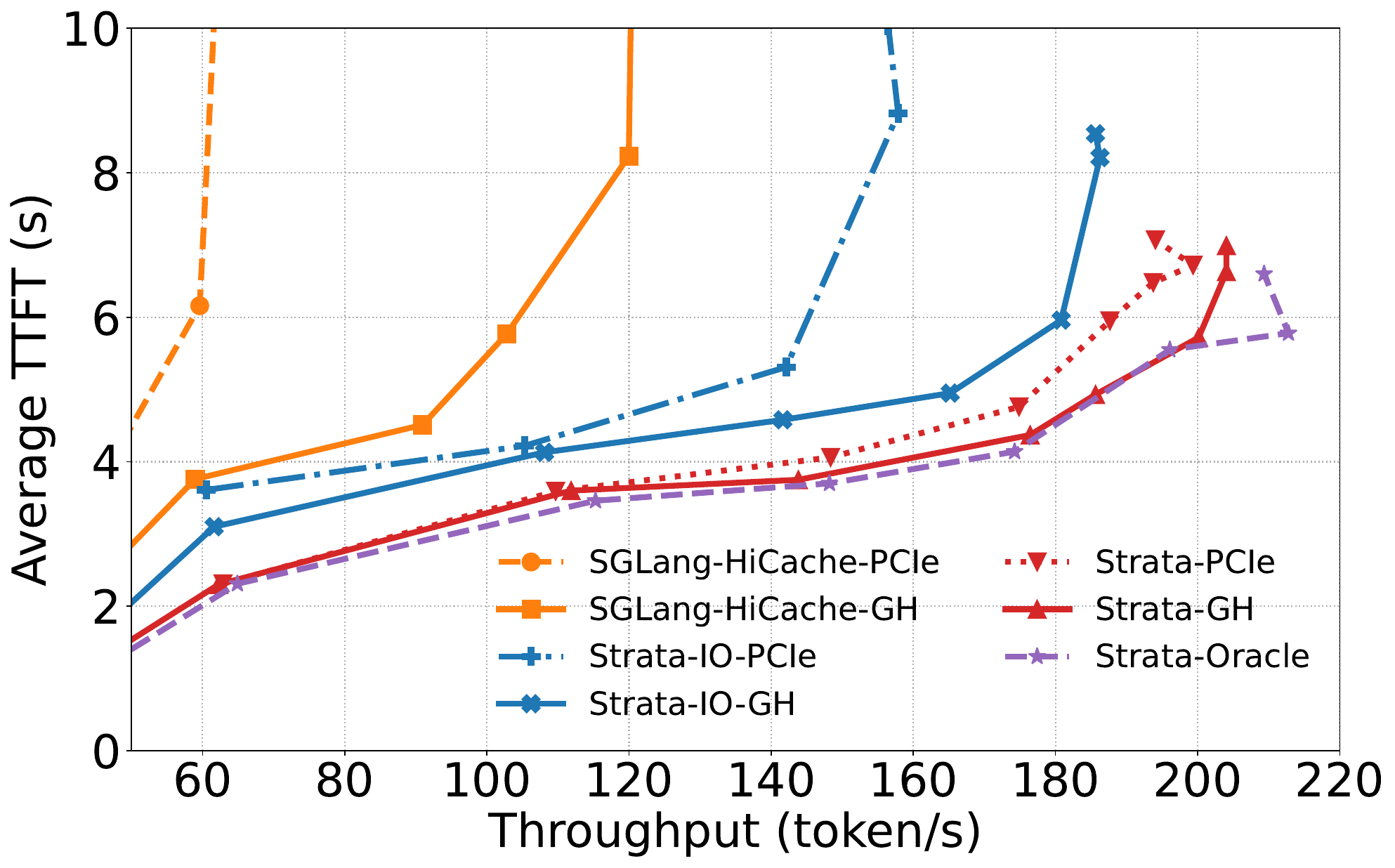}
    \caption{
    A zooming in comparison between benchmarks on PCIe-5.0 and Grace-Hopper platform.
    }
    \label{fig:grace}
\end{figure}

\begin{figure}[t]
    \centering
    \includegraphics[width=0.8\linewidth]{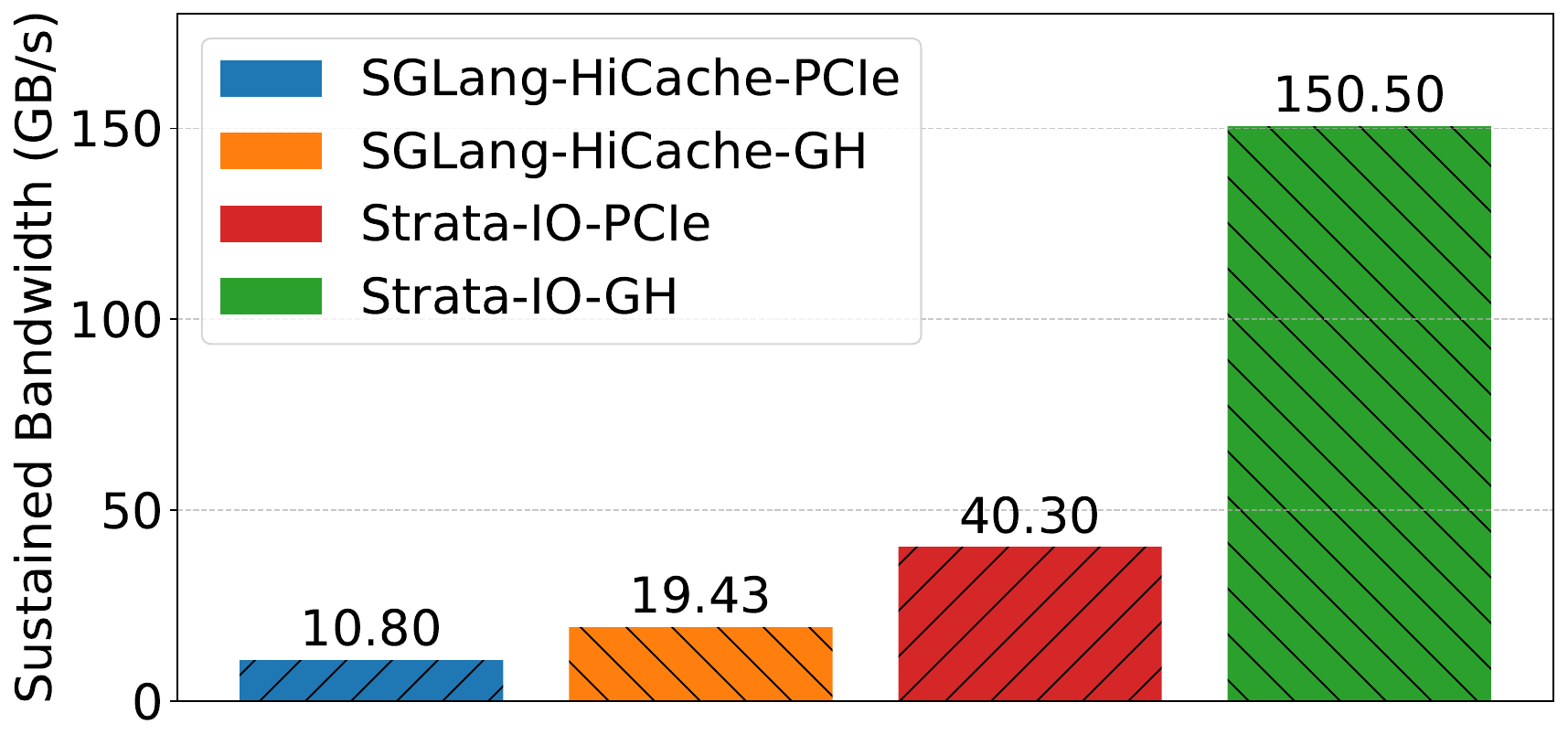}
    \caption{
    Sustained Bandwidth comparison. 
    }
    \label{fig:grace-band}
\end{figure}

Finally, we explored how \oursys aligns with emerging hardware trends that dramatically increase the bandwidth between CPU and GPU.
To do so, we benchmarked \emph{SGLang-HiCache}, \emph{\oursys-IO} and \emph{\oursys} on both our H200 and GH200 platform.
We also report an \emph{\oursys-Oracle} platform, which simulates the TTFT achieved by a system that had infinite bandwidth between the CPU and GPU.

Figure~\ref{fig:grace} report the average TTFT achieved by \textit{SGLang}, \textit{\oursys-IO}, and \textit{\oursys} across both of these platforms benchmarking Llama-3.1-8B models with LooGLE dataset.
Figure~\ref{fig:grace-band} reports the averaged sustained bandwidth for the same task.
We observe that, in line with our previous evaluation results, standard DMA-managed memory copy is not able to effectively utilize hardware improvement without using larger pages. 
While the improved bandwidth of GH200 improves latency for the SGLang baseline, only hardware improvements alone cannot outperform even \oursys-IO on the H200 platform.

In contrast, using \emph{\oursys-IO} can boost the sustained throughput from $40$ to $150$ GB/s.
However, scheduling improvements are still needed to take full advantage of platforms like Grace Hopper.
While \emph{\oursys-IO-GH} achieves higher throughput with higher bandwidth mitigating the I/O stall, it still does not even outperform \emph{\oursys-PCIe}.
In summary, \oursys's improved I/O transfer mechanisms \textit{with} its bandwidth-aware scheduler can take full advantage of emerging platforms like Grace Hopper, achieving comparable performance to the Oracle platform.
These benchmarks reveal that while increased interconnect bandwidth is beneficial, it's often under-utilized by existing software. Ideally, leveraging new hardware capabilities, as demonstrated by \emph{\oursys-GH}'s near-oracle performance, could unlock new possibilities. Specifically, this platform shows promise for specialized, cost-effective, and high-performance long-context serving.

\section{Related Work}
\noindent\textbf{Context Caching and Sharing.}
Prior studies have explored various mechanisms for reusing KV caches to eliminate redundant recomputation. For instance, SGLang~\cite{neurips24:zheng_sglang} employs a RadixTree for tracking shared context.
Other serving engines, such as vLLM~\cite{sosp23:kwon_vllm} and Mooncake~\cite{fast25:qin_mooncake}, utilize hashing mechanisms that generate unique page identifiers based on token IDs and prefix page hashes.
LMDeploy~\cite{website:lmdeploy} adopts a hybrid approach by constructing coarser-grained tries.
\oursys builds upon SGLang by extending its RadixTree to a HiRadixTree.
Other studies explore KV cache sharing beyond exact prefix contexts, exemplified by CacheGen~\cite{sigcomm25:liu_cachegen}, CacheBlend~\cite{eurosys25:yao_cacheblend}.
Unlike these approximate caching schemes,
\oursys does not impact the accuracy of requests.

\noindent\textbf{KV Cache Offloading.}
Several recent works have explored utilizing secondary memory tiers (e.g., CPU DRAM, SSDs) for KV cache storage, loading them for computation on demand. 
CachedAttention~\cite{atc24:gao_cachedattention} and Pensieve~\cite{eurosys25:yu_pensieve} both adopt a layer-wise strategy to overlap KV cache loading with computation. FlashGen~\cite{asplos25:flashgen} further enhances this pipeline with re-order execution scheduling, which has been implemented in SGLang and used in our baseline settings.

\noindent\textbf{Large-scale KV Cache Disaggregation}
Recent works propose building large-scale disaggregated KV cache memory pools and global resource coordinators to achieve caching benefits at a larger scale.
Mooncake~\cite{fast25:qin_mooncake} exploits using resources including CPU, DRAM, SSD and NIC to establish a disaggregated KV
Cache. MemServe~\cite{arxiv24:hu_memserve} unifies inter-request and intra-request KV cache optimizations via a global scheduler and an elastic memory pool.
\oursys can benefit from these designs by integrating with their KV cache transfer engines. 
However, \oursys focuses on memory management and scheduling within single compute instances and does not inherently rely on specialized hardware, such as high-speed networking, to realize its caching benefits. 
\section{Conclusion}
This paper introduced \oursys, a hierarchical context caching framework that addresses the key bottlenecks of long-context LLM serving. By combining GPU-assisted I/O to mitigate KV cache fragmentation with cache-aware scheduling to balance computation and data transfer, \oursys enables efficient utilization of system resources across diverse latency budgets. Our evaluation demonstrates that \oursys consistently outperforms state-of-the-art systems on long-context benchmarks, while maintaining strong performance on short-context workloads. These results establish \oursys as a practical and scalable solution for efficient long-context LLM serving. Looking ahead, we plan to further reduce the overhead of GPU-assisted kernels, motivating the design of more versatile on-chip memory I/O accelerators.

\begin{acks}
We thank Lingfan Yu for insightful discussions on Pensieve.
We are grateful to NVIDIA and Nebius for providing computational resources.
We also thank Sicheng Pan, Tingwei Huang, Zhangheng Huang, Arnav Balyan, Zhenwei Pi, Yi Zhang, Shangming Cai, Shiyang Chen, Ke Yang and Ying Sheng from the SGLang open-source community for their valuable feedback and contributions. 
\end{acks}

%%
%% The next two lines define the bibliography style to be used, and
%% the bibliography file.
\bibliographystyle{ACM-Reference-Format}
\bibliography{reference}

\end{document}